# A Distributed Diffusion Kalman Filter In Multitask Networks

Ijeoma Amuche Chikwendu *, Kulevome Delanyo Kwame Bensah, Chiagoziem Chima Ukwuoma, Chukwuebuka Joseph Ejiyi

University of Electronic Science and Technology of China, Chengdu (China)

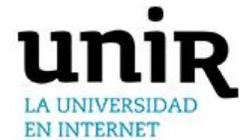

## ABSTRACT

The Distributed Diffusion Kalman Filter (DDKF) algorithm in all its magnitude has earned great attention lately and has shown an elaborate way to address the issue of distributed optimization over networks. Estimation and tracking of a single state vector collectively by nodes have been the point of focus. In reality, however, there are several multi-task-oriented issues where the optimal state vector for each node may not be the same. Its objective is to know many related tasks simultaneously, rather than the typical single-task problems. This work considers sensor networks for distributed multi-task tracking in which individual nodes communicate with its immediate nodes. A diffusion-based distributed multi-task tracking algorithm is developed. This is done by implementing an unsupervised adaptive clustering process, which aids nodes in forming clusters and collaborating on tasks. For distributed target tracking, an adaptive clustering approach, which gives agents the ability to identify and select through adaptive adjustments of combination weights nodes who to collaborate with and who not to in order to estimate the common state vector. This gave rise to an effective level of cooperation for improving state vector estimation accuracy, especially in cases where a cluster's background experience is unknown. To demonstrate the efficiency of our algorithm, computer simulations were conducted. Comparison has been carried out for the Diffusion Kalman Filter multitask with respect to the Adapt then combine (ATC) diffusion schemes utilizing both static and adaptive combination weights. Results showed that the ATC diffusion schemes algorithm has great performance with the adaptive combiners as compared to static combiners.

## KEYWORDS

Distributed Kalman filtering, Diffusion scheme, Adaptive clustering, Multi-task sensor networks .

## I. INTRODUCTION

Distributed target tracking in Wireless Sensor Networks(WSNs) remains a valuable task for various applications where a central unit is not functional. A primary goal to fix the main problems in sensor networks focuses on distributed target tracking, to deliver a real-time and accurate estimation of the target's locomotion statistics, such as location, velocity, and acceleration, on each sensor node based on not only the sensor's local measurement but incorporating shared knowledge from immediate nodes within its range of communication. Across the last decade, the subject of distributed target tracking through sensor networks has gotten a lot of attention [1], [2].

Distributed processing significantly reduces the compute load borne via the fusion center in a centralized way. Additionally, placing a vast number of sensor nodes across the surveillance zone provides superfluity in a distributed sensor network, making the overall system more resilient to sensor failures.

Distributed target tracking means that the different network nodes can collectively perform a distributed estimation in an area from their measurements. Different solutions (strategies) for solving distributed problems have been researched.

However, the diffusion-based strategies show better performance against other strategies among the available approaches, e.g., Incremental strategies [3], [4] (concerning connection robustness and hardware failures) and consensus methods [5], [6], [7] concentrating on a specific time range (in relations to permanence, convergence frequency, and steady-state performance). Recently, adaptive diffusion strategies have become the most adopted as they proffer robust solutions in distributive implementation. Initially, diffusion methods have been employed to overcome the challenges of distributed learning and adaptation [8]. They have illustrated several patterns of distributed estimating issues over networks. In [9], a distributed event-triggered estimation through a sensor network has been proposed, an energy-aware facet of distributed estimation was looked at and which is very effective at reducing unwarranted sensor samplings/transmissions, and, as a result also reduces the consumption of resource similarly to sensor power and network transmission capacity.

In [10], the Distributed Kalman Filter (DKF) was developed for linear dynamic state-space models. For sequential estimation over time, several other diffusion algorithms have been investigated [11], [12], [13]. For example, in [11], a diffusion-based approach aimed at the distributed estimate of Markov jump systems and tracking moving targets were presented, and it was used to solve a single-target tracking issue. The performance of the diffusion methods has been investigated in various scenarios [14]. An estimate-based DKF [15] was proposed and used to solve the issue of target tracking in that it relies on estimate exchange in each neighborhood. Four steps are utilized in this algorithm ranging from an individual update, local update, diffusion update, and time update for effective tracking. The authors of [16] present a finite-time distributed Kalman filter(FT-DKF) for calculating the total of universal measurement information in definite contact cycles utilizing information about measurements being diffused.

The work discussed in [10] was the initial work to suggest a diffusion-based distributed estimation fusion filter. The choice of convex combination coefficients affects the efficiency of diffusion-based algorithms. As a result, the authors discuss the best convex combination coefficients to frame a confined optimization issue in [17]. The cost-effective diffusion Kalman filter (CE-DKF) is proposed in [18], which improves efficiency by diffusing the message about state estimates and covariance estimates. The research discussed in [19] emerged as initial research using the covariance intersection (CI) approach to solve the diffusion-based distributed Kalman filtering issue and resolve the issue of sensor noise correlation. The authors projected an innovative distributed Kalman filter in [20] that uses maximum posterior probability state estimation to prevent raw data diffusion and preserve estimation accuracy.

* Corresponding author:
E-mail address: ijeomaamuche@std.uestc.edu.cn

The authors in [21] developed the non-repeated diffusion technique, which entails deleting messages obtained previously from a node and then transmitting the relevant messages to that node to prevent the information from being diffused. Even in an undirected graph, this technique causes information to diffuse only in one direction. Secondly, they applied the non-repeated diffusion approach to the diffusion-based distributed Kalman filter and proposed a novel distributed Kalman filter that uses covariance intersection to obtain convex combination coefficients.

In as much as Single-task problem has been focused on these days, more problems of interest appear to be multi-tasked which means that it has many optimum parameters or state vectors that must be inferred together collaboratively. Concerning Kalman filters, much attention has not been paid to this aspect. The main contributions of our work are as follows:

1. A derivation of a distributed algorithm for diffusion multitask in the Kalman filter domain.

2. An unsupervised clustering strategy is integrated, so network nodes can determine and select neighbors to collaborate with and improve estimation accuracy through the combination of weights adaptive adjustments method.

3. A comparative study of the diffusion multi-task performance with reverence to the static and adaptive combination weights in the Adapt then combine (ATC) diffusion scheme.

The other sections of this article are ordered as follows: the related works of multitasking are deliberated on in Section II, and the multitask problem is formulated and discussed in section III. The simulation results and discussion were the focus of section IV, and the conclusion of the paper is given in section V.

## II. RELATED WORKS

Quite numerous works have considered problems relating to multi-task scenarios. Looking at a case where different groups of network nodes track different moving targets, nodes inside the same cluster will work together to evaluate the same state vector (maybe a vector describing position or location of the target). If the targets happen to move in a similar form, then it will be beneficial for cooperation among clusters since their location vectors are related to each other. Concentrating on the distributed estimation context, many applications exist in which nodes within a network are subject to measure different model data or sensor data that vary over the spatial region. Distributed multi-sensor multitarget tracking has been researched. In [22], Üney et al. developed a distributed fusion of Probability Hypothesis Density (PHD), Cardinalized PHD (CPHD), and Bernoulli filters through merging an expanded form of covariance intersection. In [23], Battistelli et al. devised a distributed consensus CPHD filter in which individual nodes compute their local estimate based on its evaluation and request consensus iterations to attain universal fusion across the network by iteratively fostering neighborly fusion. Lately, for multitarget monitoring, a distributed particle filter enactment of the PHD filter has been proposed by Leonard et al. [24]. This algorithm has immense computational intricacy and transmission load because many weighted particles are created at individual nodes and disclosed amid neighbors for the phases of adaptation and combination. Other multitask scenarios are researched in the Diffusion Least Mean Square and Recursive Least Square domain [25], [26], [27], [28]. In Diffusion Kalman filter, little or no work has been carried out regarding multitasking. Distributed Kalman filtering is among the essential information processing procedures in WSNs.

Owing to its elemental state-space model, which considers observational noise, it has been verified to be favorable in relation to heightened precision and speedy convergence range. In light of this, this work applies the multi-task to the distributed DKF. The Kalman Filter (KF) is better suited for applications that may need fast-tracking and precise tracking of the unknowns, especially if the devices can deal with computations that are moderately high in complexity compared to the Least Mean Square (LMS); this motivates the researcher to research the distributed multitask learning DKF problem or networks. The Kalman filter algorithm would be focused on as it can be seen as the workhouse for tracking in the sense that it is an optimal minimum mean squared error (MMSE) estimator. It is a (recursive) weighted sum of the prediction and observation.

In other words, the Kalman filter is a context for predicting a process state while using measurements to correct or "update" the predictions. In this multi-task sensor network, which is used to track numerous state vector targets [29], fusing information across nodes tasked with distinct objectives may compromise their effectiveness and provide outcomes that may be undesired and spread through the network [30]. Currently, adaptive real-time approaches that can group nodes of a network monitoring a shared objective are in great demand. Most adaptive clustering techniques have been designed using distributed least mean square procedure as their reference [31], [32], [33].

Even though these techniques can be useful to the network sensors of the distributed DKF domain, they don't utilize the extra information provided in the KF like the state evolution model and the state vector estimate error covariance matrices. To this end, in this work, a distributed DKF algorithm for a multi-task network is derived, drawing ideas from [34]. The derivation gives awareness to the process of the distributed DKF, permitting an adaptive clustering technique to be established. The adaptive clustering method utilizes the covariance information present in the Kalman filtering procedure to recursively revise the state estimates from one network node, thereby making available valid information regarding the state vector information of its neighbors. To the best of our knowledge, there is not much work on multi-tasking for the Kalman filter, probably because of the fundamental challenges of target tracking. So, our idea is to propose a distributed algorithm for multi-tasking in the Kalman filter domain.

### A. Adaptive Combination Weights

The weights of each neighbor's combination play a significant role in the success of adaptive networks [35], [36]. The Uniform rule [37], Metropolis rule [38], and Relative-degree rule [39] are some proposed combination policies in previous studies. To assign combination weights, they bank primarily on the degree of the nodes. Since they disregard the noise profile across a network, these choices can degrade adaptive network performance. Since nodes with more neighbors may have a lower signal-to-noise ratio than nodes with fewer neighbors, constructing combination weights for their neighbors solely based on the nodes' degrees is inadequate. As a result, it is essential to think about the noise profile in the nodes while designing the combination strategy. Some earlier work in this area includes [40], [41]. In [40], some scholars suggested a combination rule called relative degree-variance taking into consideration the noise profile of nodes. It is observed that [41], with their adaptive combination rules has achieved reduced steady-state error at a substantial penalty to convergence speed. Other combination schemes are implemented [42], [43]. [44] based on the purpose of reducing the mean square deviation (MSD) of the proposed diffusion adaptive networks, two effective adaptive combination strategies called relative-instantaneous-error combination strategy and relative-deviation combination strategy are associated with the opposite of noise by diverse metrics.

**Mathematical Notation**: Boldface lowercase letters symbolize vectors and boldface uppercase letters symbolize matrices. The superscripts $(.)^T$ and $(.)^*$ signify transpose and the complex conjugate transpose of a matrix/vector, correspondingly. The notation $col\{.\}$ represents the vector attained by stacking its entries in succession to one another. Similarly, we use $diag\{.\}$ to represent the (block) diagonal matrix comprising the specified vectors or matrices. The mathematical expectation is expressed by $E\{.\}$. $\{.\}^{-1}$ represents the matrix inverse. The operator $||.||$ denotes the l$_2$-norm of a vector. Normal font letters symbolize scalars. $\mathbf{I}_N$ Identity matrix of size $N \times N$. $N_m$ represents the index set of nodes in the node's neighborhood $m$, including $m$. $N_m^-$ symbolizes the index set of nodes in the node's neighborhood $m$, exclusive of $m$. $C_i$ represents cluster $i$, i.e., index set of nodes in the $i-th$ cluster. $C(m)$ represents the cluster of that node k is part of, i.e.,$C(m) = \{C_i : m \in C_i\}$.

## III. MULTITASK PROBLEMS AND DIFFUSION KALMAN FILTER

The local optima differ amongst clusters in a multitask network. By lessening the cluster cost function, which combines individual nodes' local cost functions from equivalent clusters in a distributed fashion, nodes belonging to an equivalent cluster can achieve a shared optimum. Nodes just swap information locally and collaborate with their sub-neighbors in the same cluster, with no requirement to share or need task-irrelevant universal information.

### A. Network with Various Clusters

A linked network having a node-set $S = \{1, 2, \dots N\}$ which is classified into $s$ commonly unique clusters, indicated by $\{C_l\}_{l=1}^s$ is studied. Each network node $m$ can convey information with its bordering agents $N_m$. A real-time cluster containing $m$ is characterized by $C_{m,j} = N_m \cap N_{m,j}^+$, where $N_{m,j}^+$ portrays sub neighbors having equivalent objectives as node $m$ acquired by clustering detection at time $j$. $N_{m,j}^- \triangleq N_m \backslash C_{m,j}$ depicts sub neighbors with the disparate objective with that of node $m$ at time $j$. However, the sub neighbors in equivalent or disparate clusters of nodes in the initial stage remain fuzzy. Figure 1 depicts 10 nodes divided into two clusters, with colored circles illustrating nodes within distinct clusters.

The solid lines connect sub-neighbors in an equivalent cluster at either end, whereas the dotted lines connect sub-neighbors in separate clusters. Categorically, in Figure 1(a), node $m$ and its sub neighbors of equivalent cluster form $N_{m,j}^+ = \{n, m, 3,4,5,6\}$ whereas the sub neighbors of a cluster distinct from node $m$ form $N_{m,j}^- = \{7\}$. The clusters emerge with time concerning the dynamic clustering detection utilized in multitask networks. Figure 1(b) depicts the steady-state subnetworks for every cluster after the transformative clustering process.

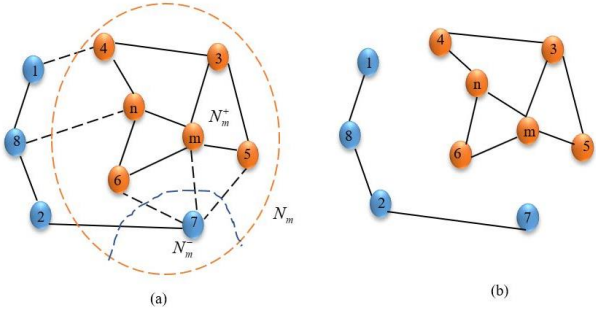

Fig. 1. Network topology for multitasking: (a) networks with two clusters and (b) network clusters after the transformative clustering process.

### B. Problem Formulation

Suppose that the sensor network has $N$ sensor nodes and there are total $N_t$ targets, where $N_t \geq 1$. We focus on the 2D tracking scenario. The state vector of target $i$, where $i = 1, \dots, N_t$, time index j is $X_{i,j} = [x_{i,j}, y_{i,j}, \dot{x}_{i,j}, \dot{y}_{i,j}]^T$ where $[x_{i,j}, y_{i,j}]^T$ and $[\dot{x}_{i,j}, \dot{y}_{i,j}]^T$ are the position and velocity of the $i^{th}$ target at time index j. The evolution of the target state vector is modeled as follows in Equation (1).

$$\mathbf{X}_{i,j} = \mathbf{F}_{i,j}\mathbf{X}_{i,j} + \mathbf{G}_{i,j}\mathbf{u}_{i,j}\mathbf{w}_{i,j} \quad (1)$$

Where $\mathbf{F}_{i,j}$ is the transition matrix and $\mathbf{w}_{i,j}$ is the Gaussian system noise with zero mean and covariance $\mathbf{Q}_{i,j}$. The measurement vector $\mathbf{y}_{m,i,j}$ is related to the target state vector $\mathbf{x}_{i,j}$ through

$$\mathbf{y}_{m,i,j} = \mathbf{H}_{m,i,j}X_{i,j} + \mathbf{v}_{m,i,j} \quad (2)$$

Where $\mathbf{H}_{m,i,j}$ and $\mathbf{v}_{m,i,j}$ are the measurement matrix and measurement noise of node $m$ of the $i^{th}$ target at time $j$ respectively.

A multitask network environment is studied in which distinct clusters execute different tasks. Each agent $m$ has an interest in monitoring and tracking a unique M × 1 unidentified optimum state vector $\mathbf{x}_m^0$. Nodes of equivalent clusters estimate the same optimum vector

$$\mathbf{x}_m^0 = \mathbf{x}_{C_l}^0 \forall m \in C_l \quad (3)$$

where the cluster $C_i \in \{C_1, C_2, \dots C_s\}$ represents the multi-task network and every node $m$ gathers measurement during time $j$ for the $i^{th}$ target as indicated in Equation (2). It is possible to write a single node cost function that minimizes the overall cost function is the mean squared deviation form as shown in Equation (4)

$$MSD_{m,j|k} = \sum_{l=1}^{s} \sum_{m \in C_l} E||\mathbf{x}_j - \hat{\mathbf{x}}_{m,j|k}||^2 \quad (4)$$

for the cluster $C_i \in \{C_1, C_2, \dots C_s\}$.

### C. Adaptive Clustering Strategy

The algorithm is initialized by specifying the number of clusters to assign tasks and then randomly selecting a node to act as a cluster head. The adjacency matrix is then used to decide its neighbors based on a given radius to perform the same task as one cluster. Automatically, any node not assigned to the first cluster is then in cluster two. Since our study focuses on two tasks, all other nodes are now clustered to perform the second task. After this is done, neighboring nodes are determined, and links between inter clusters are broken. This strategy is shown in Algorithm 1.

| Algorithm 1: Adaptive clustering strategy |
|---|
| **Initialization:** |
|     Specify the number of clusters to assign |
|     # clusters = #targets/tasks |
|     $C$ = # clusters |
|     $r$ = radius to determine nodes in the $m$ |
| **Step 1:** |
|     For $C = 2$ do |
|       Perform clustering to partition the network into 2 |
|       Randomly select a node as a cluster head $(m_h)$ |
|         For $m = 1: N$ |
|           If dist $(m, m_h) \leq r$ then |
|             add a node $m$ to cluster 1 |
|           End if |
|         End for |

## D. Node Clustering By Combination Matrix Selection

To adapt to a multi-task environment, a clustering method in which individual node $m$ can alter the combination weights $C_{nm}$ in an online setting, for $n \in N_m$ is used. The strategy is achieved by adjusting the **C** matrix. Consider different cost functions for individual nodes to improve the versatility of multitask networks. This introduces the issue of information distribution through the transfer matrix **A**, which can be easily set to identity. Equation (5) depicts how each node mixes the state vectors conveyed by its neighbors in relation to the projected task contrast.

$$c_{nm}(j+1) = \frac{||\Psi_m(j+1) + q_m(j) - \Psi_n(j+1)||^{-2}}{\sum_{l \in N_m} ||\Psi_m(j+1) + q_m(j) - \Psi_l(j+1)||^{-2}} \quad (5)$$
$$for\ n \in N_m$$

where $q_m$ is an instantaneous error. This rule is useful in that it depends on the local estimate to decrease the mean square deviation (MSD) bias impact produced via the collaboration of close nodes calculating various state vectors. The combination rule, as shown in Equation (5), to adjust the combination weight, considers the local estimate's proximity to nearby estimations as well as the cost function's local slope. This promotes cooperation among nodes that estimate the same ideal parameter. As a result, the MSD bias is reduced, and the estimation accuracy is improved. Another way to make use of this information is for every agent to apply the mutual dependence precept established by

$$\mathbf{A}_{j|j+1} = \mathbf{C}_{j|j+1}^T \quad (6)$$

The reasoning behind this precept is that the value of $C_{nm}$ represents the resemblance of the tasks carried out by nodes $m$ and $n$ together with how they are interpreted by node $m$. This knowledge should be used by the node $n$, and the local cost function should be scaled accordingly. Since nodes $m$ and $n$ do not solve the same estimate problem, the smaller $C_{nm}$ is, the smaller $a_{nm}$ should be. Algorithm 2 provides an overview of the ATC diffusion method with adaptive clustering described by the time-variant combination matrices $\mathbf{C}(j)$ and $\mathbf{A}(j)$. Considering that no preceding knowledge on clusters exists, initializing the combination matrices $\mathbf{C}(0)$ and $\mathbf{A}(0)$ with $\mathbf{I}_N$ is essential.

## E. ATC Diffusion Kalman Filter (DKF) Algorithm with Adaptive Clustering for Multitask

The DKF is studied under the diffusion scheme algorithm known as Adapt then Combine (ATC).

In the ATC diffusion scheme, the nodes adapt and exchange the measurement with their neighbors and get the intermediary estimate utilized in the combination step to get the optimum state vector. Algorithm 2 is composed of two steps: incremental update and diffusion update, which can be seen as adaptation and combination respectively. Following the initialization of the combination matrices, the first adaptive phase employs an iterative approach to minimize the individual node cost function. Nodes exchange local data amongst neighbors and at each point in time $j$, updates $\Psi_{m,j} \leftarrow \hat{x}_{m,j|j-1}$ and $P_{m,j} \leftarrow P_{m,j|j-1}$. After that, every node accomplishes the Kalman filter with the data it has to generate the intermediate estimates thereby adding innovation. The next step represents a convex combination of the intermediate estimate $\Psi_{m,j}$ from the Kalman Filter which is shared by geographically separate data from $l \in N_{m,j}^+$, the sub neighbors with mutual objectives. That is, it cooperates with sub-neighbors of equivalent clusters.

---

**Algorithm 2: ATC DKF with adaptive clustering for multitask**

**Initialization:** Set $A(0) = I_N$ and $C(0) = I_N$
Set $\hat{X}_{m,0|-1} = 0$ and $P_{m,0|-1} = \Pi_0$
For every time instant $j$, every node $m$ compute
**Step1: Adaptation Phase**

$\Psi_{m,j} \leftarrow \hat{x}_{m,j|j-1}$
$P_{m,j} \leftarrow P_{m,j|j-1}$

For every neighboring node $n \in N_m$, repeat

$R_e \leftarrow R_{n,j} + H_{n,j} P_{m,j} H_{n,j}^*$
$\Psi_{m,j} \leftarrow \Psi_{m,j} + P_{m,j} H_{n,j}^* R_e^{-1} [y_{n,j} - H_{n,j}\Psi_{m,j}]$
$P_{m,j} \leftarrow P_{m,j} - P_{m,j} H_{n,j}^* R_e^{-1} H_{n,j} P_{m,j}$

end for
Update combination coefficients

$q_{m,j} = [y_{m,j} - H_{m,j}\Psi_{m,j}]$

$$c_{nm}(j+1) = \frac{||\Psi_m(j+1) + q_m(j) - \Psi_n(j+1)||^{-2}}{\sum_{l \in N_m} ||\Psi_m(j+1) + q_m(j) - \Psi_l(j+1)||^{-2}}$$

**Step 2: Combination Phase**

$$\hat{X}_{m,j|j} = \sum_{n \in N_m} c_{nm}(j+1) \Psi_{m,j+1}$$
$$P_{m,j|j} \leftarrow P_{m,j}$$
$$\hat{X}_{m,j+1|j} = F_j \hat{X}_{m,j|j}$$
$$P_{m,j+1|j} = F_j P_{m,j|j} F_j^T + G_j Q_j G_j^T$$

---

## IV. SIMULATION RESULTS AND DISCUSSIONS

### A. Experimental Setup

This segment includes simulations and performance analyses of adaptive clustering and distributed tracking. The simulation is run on a linked network with N = 30 nodes, in a randomly created topology. The challenge with estimating and monitoring the location of a projectile is considered, in which the sensors from different clusters are tasked to obtain noisy measurements of the projectile's position. To illustrate the performance of the developed DDKF clustering system works, we considered a target tracking application over the network with the main aim of each cluster to estimate the precise location of the projectile at each time instant in the Kalman filtering problem. The system's state is a two-dimensional object's unknown location vector, represented by the coordinates $(x, y)$ where $x$ and $y$ are the first and second entries, correspondingly. In the simulation example, the projectile's location, velocity, and acceleration are represented in Equation (7).

$$\mathbf{a} = \begin{bmatrix} a_x \\ a_y \end{bmatrix}, \mathbf{v} = \begin{bmatrix} v_x \\ v_y \end{bmatrix} \ \mathbf{d} = \begin{bmatrix} d_x \\ d_y \end{bmatrix} \quad (7)$$

For the projectile motion, we have Equation 8

$$\mathbf{a} = \dot{\mathbf{v}}, \mathbf{v} = \dot{\mathbf{d}} \quad a_x = 0, a_y = -g \quad (8)$$

Where $g = 10$, the acceleration due to gravity is constant. The state **X** of the system is gotten by stacking the position and velocity of the projectile. Equation (9) thus provides the following description of the state equation:

$$\underbrace{\begin{bmatrix} \dot{d} \\ \dot{v} \end{bmatrix}}_{\dot{x}} = \underbrace{\begin{bmatrix} 0 & I_2 \\ 0 & 0 \end{bmatrix}}_{\theta} \underbrace{\begin{bmatrix} d \\ v \end{bmatrix}}_{x} + \underbrace{\begin{bmatrix} 0 \\ \begin{bmatrix} 0 \\ -g \end{bmatrix} \end{bmatrix}}_{n} \quad (9)$$

This can be written in a compact form $\dot{\mathbf{x}} = \boldsymbol{\theta}\mathbf{x} + \mathbf{n}$. Note that for the matrix $\boldsymbol{\theta}$, we have

---

**Step 2:** All selected nodes = cluster 1
Remaining nodes in network = cluster 2
End for

$$e^{\theta\delta} = \mathbf{I} + \delta\theta \text{ and } \int_{j_0}^{j_0+\delta} e^{\theta(j_0+\delta-\tau)} d\tau \qquad (10)$$

As shown in Equation (10). Therefore, the state satisfies Equation (11).

$$\mathbf{x}(j+\delta) = [\mathbf{I} + \delta\theta]\mathbf{x}(j) + [\delta\mathbf{I} - \delta^2\theta/2]\mathbf{n} \qquad (11)$$

In Equation (12), for a specific time step $\delta$, **F**, and **u** can be represented as follows

$$\mathbf{F} \triangleq \mathbf{I} + \delta\theta \text{ and } \mathbf{u} \triangleq [\delta\mathbf{I} - \delta^2\theta/2]\mathbf{n} \qquad (12)$$

It is presumed that every cluster node calculates the location and velocity of the unfamiliar target in the two dimensions, i.e., x and y.

The experimental results are achieved by collectively averaging over 200 independent trials. Assuming the moving target's initial projectile position is $x_0 = 1$, $y_0 = 30$ and the initial velocity for both targets is $v = 15$, $\theta = \pi/3$ and $\theta = \pi/4$ is the angle of targets 1 and 2 respectively. $v_{x_0} = v * \cos(\theta)$, $v_{y_0} = v * \sin(\theta)$ represents the x and y speed. The values of the rest of the parameters used in the algorithm are as follows. The size of the state vector, $M = 4$, measurement noise covariance matrix at the node $m$ is $\mathbf{R}_{m,j} = \sigma_{m,j}^2 \mathbf{I}_4$ where the noise variance $\sigma_{m,j}^2$ across the nodes is arbitrarily gotten in the space 0.5*rand(N,1)+0.01, where rand(N,1) represents the arbitrary number in the space of 0 to 1. The first value of the state covariance matrix is $\mathbf{P}_0 = \mathbf{I}_0$, State noise matrix $\mathbf{G}_j = 0.625\mathbf{I}_4$, Covariance state noise matrix $\mathbf{Q}_j = 0.001\mathbf{I}_4$.

Observation matrix: $\mathbf{H}_{m,j} = \begin{bmatrix} 1 & 0 & 0 & 0 \\ 0 & 1 & 0 & 0 \\ 0 & 0 & 1 & 0 \\ 0 & 0 & 0 & 1 \end{bmatrix}$

State transition matrix: $\mathbf{F}_j = \begin{bmatrix} 1 & 0 & T & 0 \\ 0 & 1 & 0 & T \\ 0 & 0 & 1 & 0 \\ 0 & 0 & 0 & 1 \end{bmatrix}$

### B. Node clustering and Trajectory Tracking

In the considered linked network, all nodes have neighbors and none is isolated. Each node of this network has not less than four single-step reachable neighbors and at least a neighbor who belongs to an equivalent cluster.

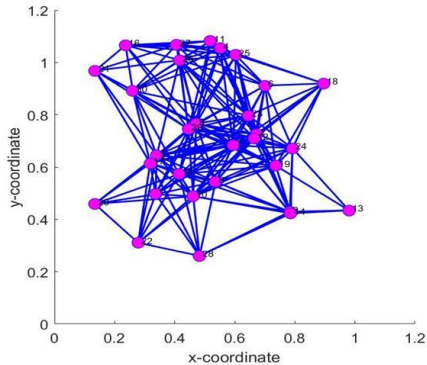

Fig. 2. Initial global network topology

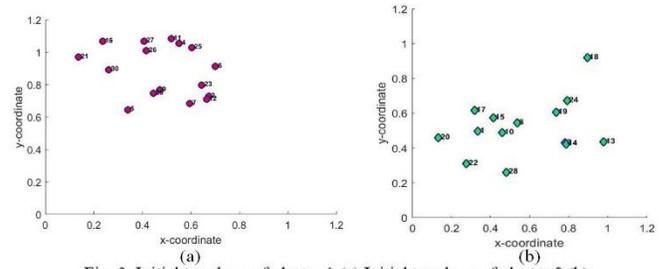

Fig. 3. Initial topology of cluster 1 (a) Initial topology of cluster 2 (b)

It is presumed that each node measures both position and velocity of the unknown target in the two dimensions, x, and y.

The 30-node network is divided into two mutually exclusive clusters with cluster $C_1 \epsilon \{21,16,27,26,11,4,25,6,23,2,7,12,29,9,5,30\}$ node corresponding to the color red and green represents cluster $C_2 \epsilon \{20,22,28,1,17,15,10,8,19,24,18,13,14,3\}$ nodes respectively. Figure 2 depicts the initial network topology at the first stage with all potential linkages depicted.

As illustrated in Figures 3 (a) and (b), nodes in other clusters do not initially create cluster structures because it is unknown if a sub neighbor is in an equivalent or separate cluster of nodes, and nodes have no idea of its cluster at the start.

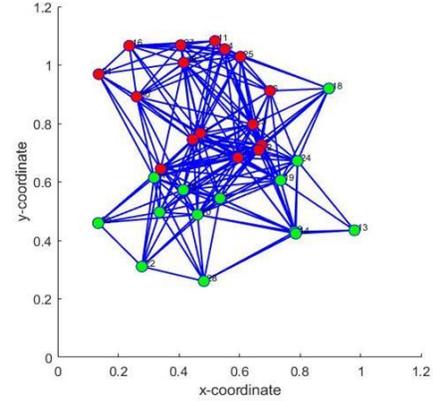

Fig. 4. Adaptive clustering topology

After the clustering operation, the stable network architecture that includes subnetworks for each cluster and a clustered structure is illustrated in Figure 4. Each cluster is produced once all the relationships between sub-neighbors in the distinct clusters are separated, as seen in Figures 5 (a) and (b).

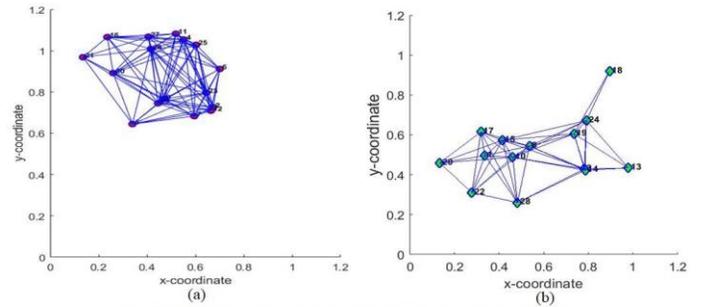

Fig. 5. (a) Final topology of cluster 1 and (b) Final topology of cluster 2

The simulation compares the results of the three static combination weights uniform, metropolis, and relative variance, to the adaptive combination weights diffusion ATC scheme. However, in as much as the clusters of the network have been able to track their various task, it is noticed that the adaptive weights have better performance compared to the static weights of which they are superimposed. This is because with respect to time, networks can track the trajectory of the projectile and how the different weight behaves.

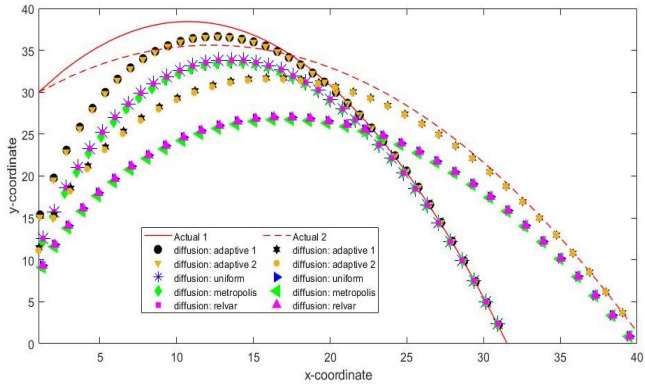

Fig. 6 Trajectory plot showing the performance of the two clusters with the various combination rules.

## C. Global Mean-Square Performance

Mean-Square Deviation (MSD) is an important parameter for evaluating an algorithm's error convergence ability. MSD primarily replicates the difference between the real state $\mathbf{X}_{i,j}$ and estimate state $\hat{\mathbf{X}}_{i,j}$. In Figure 9, the MSD of the algorithms over the entire network is displayed. The ATC algorithm converges after 40 iterations. Fast convergence is recorded proving the policy apposite for a dynamic real-time system. The following conclusion is drawn through the algorithm result: the diffusion strategies are more fit to distributed algorithms, resulting from the weight value $C$ denoted in Algorithm 2. The $C$ value is left stochastic and is more appropriate for the characteristics of the distributed network.

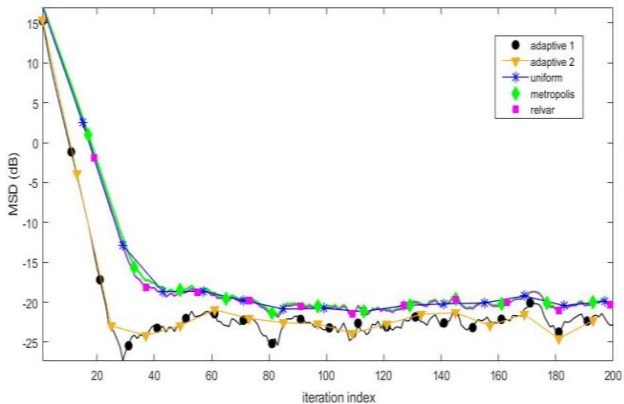

Fig. 7 MSD plot for cluster 1.

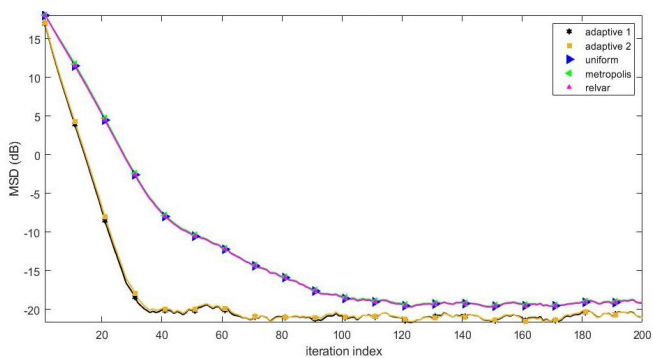

Fig. 8 MSD plot for cluster 2.

At the first 40 iterations, the clusters through the Kalman diffusion strategy get used to their task and eventually can track their projectile. It is seen from the MSD plots in Figures 7 and 8 that adaptive weights have outperformed the static combination weights. it is also observed that there is no significant difference between the static combination weights. The adaptive combiners can adjust and assign a weight to befitting neighbors. Figure 6 shows the plot for the two clusters.

## V. CONCLUSIONS

This study focused on distributed diffusion Kalman filters and clustering framework for multi-task networks through adaptive clustering. An adaptive clustering method that utilizes an adjustment via adaptive combination weights for accurate clustering was used, which enabled nodes to select and collaborate with nodes within their cluster. The ATC diffusion scheme was implemented in the multi-task problem, and a comparison was made between static combination and adaptive rules to ascertain its performance. Only the exchange of local state vector estimations and their related covariance information is vital for the established framework. We have carried out simulations and presented simulation results illustrating the performance of the diffusion Kalman filter multi-task algorithms, which shows the algorithm was able to form clusters adaptively and track the target. The limitations of this work are:
1.) The proposed study focused on tracking a minimum of two objects due to computational complexities.
2.) More so, with respect to time constraints, the study only focused on a stationary environment with a static number of targets at every time instant.

### *Future work*

The scope of this study focused on two targets and hence two clusters were generated adaptively. However, to extend the scope of this study, there is room to improve and explore the algorithm studied in this work by considering more targets which will mean segregating the network adaptively to track these targets.

Energy efficiency is another fascinating issue that analyzes how to reduce the energy required for sensor motions.

Also, implementing the algorithm in an environment that is non-stationary and where the number of targets is not constant, changing at a specific point in time would be an interesting research area.

## ACKNOWLEDGMENT

This work was partly supported by the National Natural Science Foundation of China under Grants 61571099, and 61501098.

## DECLARATION OF INTEREST

We the authors declare that there is no conflict of interest.

diffusion adaptive networks," *Peer-to-Peer Netw. Appl.*, vol. 13, no. 1, pp. 123–126, 2020, doi: 10.1007/s12083-019-00726-2.

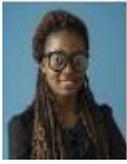

### Ijeoma Amuche Chikwendu

Ijeoma Amuche Chikwendu Received a B.Sc degree in Information management technology at the Federal University of Technology Owerri in 2014 and a Masters degree in Information and Communication Engineering at the University of Electronic Science and Technology of China (UESTC) in 2021. She is currently pursuing a Ph.D. degree at the same University where she obtained her master's degree. Her research interest is Distributed estimation and target tracking, and currently working on Graph representation learning and deep learning.

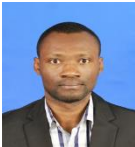

### Kulevome Delanyo Kwame Bensah

Kulevome Delanyo Kwame Bensah was born in 1983. He received his B. Eng. degree from Accra Institute of Technology, and M. Eng. degree in Electronic Science and Technology in 2019 from the University of Electronic Science and Technology of China, Chengdu, China, where he is currently pursuing his Ph.D. degree in Information and Communication Engineering. His research interests include prognostics and health management of systems, fault diagnostics, signal processing, and deep learning.

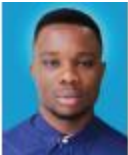

### Chiagoziem C. Ukwuoma

Chiagoziem C. Ukwuoma received his B.Eng. degree (Mechanical Engineering-Automobile Technology) from the Federal University of Technology Owerri in 2014 and his MSc. degree (Software Engineering) from the University of Electronic Science and Technology of China (UESTC) in 2020. He is currently a Ph.D. student at the University of Electronic Science and Technology of China (UESTC). His research interests include Object Detection and Object Classification.

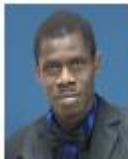

### Chukwuebuka Joseph Ejiyi

Chukwuebuka Joseph Ejiyi received his Bachelor's Degree in 2014 from the Federal University of Technology Owerri (FUTO) Nigeria. He went on to obtain a master's degree in Software Engineering at the University of Electronic Science and Technology of China (UESTC) in 2021. He is currently pursuing a Ph.D. degree with the Schoool of Information and Software Engineering at UESTC Chengdu China. His research interest is in Artificial intelligence, Deep Learning and he is currently working on Object detection using a single-stage neural network as well as Object classification. He also has a strong interest in image analysis, especially with regard to medical images.